\documentclass{article}  
\usepackage{authblk}
\usepackage{graphicx,natbib}
\usepackage{amsmath,amssymb}
\usepackage{bbm}
\usepackage{tablefootnote}
\usepackage[pagebackref=true,colorlinks,linkcolor=blue,citecolor=magenta]{hyperref} 
\def \RR{{\mathbb R}}

\def \EE{{\mathbb E}}
\def \PP{{\mathbb P}}
\def \Cov{{\mathbb Cov}}
\def \Var{{\mathbb{V}\mathrm{ar}}}

\newcommand{\I}{{\mathbbm{1}}}

\begin{document}

\title{A hierarchical spatio-temporal model to analyze relative risk variations of COVID-19: a focus on Spain, Italy and Germany
}

\author[1]{Abdollah Jalilian}
\author[2]{Jorge Mateu}
\affil[1]{Department of Statistics, Razi University, Kermanshah, 67149-67346, Iran}
\affil[2]{Universitat Jaume I, Castell\'{o}n, Spain}



\date{}

\maketitle

\begin{abstract}
The novel coronavirus disease (COVID-19) has spread rapidly across the world in a short period of time and with a heterogeneous pattern. Understanding the underlying temporal and spatial dynamics in the spread of COVID-19 can result in informed and timely public health policies. In this paper, we use a spatio-temporal stochastic model to explain the temporal and spatial variations in the daily number of new confirmed cases in Spain, Italy and Germany from late February to mid September 2020. Using a hierarchical Bayesian framework, we found that the temporal trend of the epidemic in the three countries rapidly reached their peaks and slowly started to decline at the beginning of April and then increased and reached their second maximum in August. However decline and increase of the temporal trend seems to be sharper in Spain and smoother in Germany. The spatial heterogeneity of the relative risk of COVID-19 in Spain is also more pronounced than Italy and Germany.


\end{abstract}

\section{Introduction}
\label{sec:intro}

Started from Wuhan, the capital of Hubei province, China in December 2019, the outbreak of 2019 novel coronavirus disease (COVID-19) has spread rapidly across more than 200 countries, areas or territories in a short period of time with so far over 4.4 million confirmed cases and 296 thousand confirmed deaths
\citep{world2020coronavirus}. 

The spread of COVID-19 across and within countries has not followed a homogeneous pattern \citep{giuliani2020modelling}.  
The causes of this heterogeneity are not  yet  clearly  identified, but different countries have different  levels of national capacity based on their abilities in prevention, detection, response strategies, enabling function, and operational readiness \citep{kandel2020health}. Besides, different countries have implemented different levels of rigorous quarantine and control measures to prevent and contain the epidemic,  which affect the population movement and  hence the spread pattern of COVID-19. 
Given the highly contagious nature of COVID-19, the spatial pattern of the spread of the disease  changes rapidly over time. Thus, understanding the spatio-temporal dynamics of the spread of COVID-19 in different countries is undoubtedly critical. 

The spatial or geographical distribution of relative location of incidence (new cases) of COVID-19 in a country is important in the analyses of the disease risk across the country. 
In disease mapping studies, the spatial domain of interest is partitioned into a number of contiguous smaller areas, usually defined by administrative divisions such as provinces, counties, municipalities, towns or census tracts, 
and the aim of the study is to estimate the relative risk of each area at different times \citep{lee2011comparison,lawson2018book}. Spatio-temporal models are then required to explain and predict the evolution of incidence and risk of the disease in both space and time simultaneously \citep{anderson2017comparison}.

Estimation of area-specific risks over time provides information on the disease burden in specific areas and identifies areas with elevated risk levels (hot spots). In addition, identifying the changes in the spatial patterns of the disease risk over time
 may result in detecting either regional
or global trends, and contributes to make informed and timely public health resource allocation \citep{wakefield2007disease}.

To account for the underlying temporal and spatial autocorrelation structure in the spread of COVID-19, available data on the daily number of new cases and deaths in different countries/regions have already been analyzed in a considerable number of studies. 
For example, \cite{kang2020spatial} used  Moran's I spatial
statistic with various definitions of neighbors and observed a significant spatial association of COVID-19 in daily number of new cases in provinces of mainland China. 
\cite{gayawan2020spatio} used a zero-inflated Poisson model for the daily number of new COVID-19 cases  in the     African continent and found that the pandemic  varies 
geographically across Africa with notable  high incidence in neighboring countries. 
\cite{briz2020spatio} conducted a spatio-temporal analysis for exploring the effect of daily temperature on the accumulated number of COVID-19 cases in the provinces of Spain. They found no evidence suggesting a relationship between the temperature and the prevalence of COVID-19 in Spain. 
\cite{gross2020spatiotemporal} studied the spatio-temporal spread of COVID-19 in China and compare it to  other  global  regions and concluded that human mobility/migration from Hubei and the spread of COVID-19  are  highly  related. 
\cite{danon2020spatial} combined 2011 census data to capture population sizes and population movement in England and Wales with parameter estimates from the outbreak in China and found that the COVID-19 outbreak is going to peak around 4 months after the start of person-to-person transmission. 
Using linear regression, multilayer perceptron and vector autoregression, \citet{Sujath:2020} modeled and forecasted the spread of COVID-19 cases in India.


As pointed out in \cite{alamo2020open}, there are many national and international organizations that provide open data on the number of confirmed cases and deaths. However,  these data often suffer from incompleteness and inaccuracy, which are considerable limitations for any analyses and modeling conducted on the available data on COVID-19 \citep{Langousis:2020}.
We highlight that we are yet in the center of the pandemic crisis and due to the public health problem, and also to the severe economical situation, we do not have access to all sources of data. Thus reseachers know only a portion of all the elements related to COVID-19. 
In addition, data on many relevant  variables such as population movement and interaction and the impact of quarantine and social distancing policies are  not either available or  accurately  measured. Combined  with the unknown  nature of the new COVID-19 virus, any analysis such as the present study only provides an approximate and imprecise description of the underlying spatio-temporal dynamic of the pandemic. Nevertheless,  having a vague idea is better than having no idea, and the results should be interpreted with caution. 

Currently, a wealth of studies have appeared in the very recent literature. Many of them follow the compartmental models in epidemiology, partition the population into subpopulations (compartments) of susceptible (S), exposed (E), infectious (I) and recovered (R), and fit several variations of the classical deterministic SIR and SEIR epidemiological models 
\citep{peng2020epidemic,roda2020difficult,bastos2020modeling}. 
We believe that considering stochastic components is important, if not essential, to explain the complexity and heterogeneity of the spread of COVID-19 over time and space. For this reason, in the present work  we propose a spatio-temporal stochastic modeling approach that is able to account for the spatial, temporal and interactions effects, together with possible deterministic covariates.

We acknowledge that the proposed model in its current form requires development and refinements  as more information becomes available, but at the stage of the pandemic we are now, it can provide a reasonable modeling framework for the spatio-temporal spread of COVID-19. This is illustrated by modeling the daily number of new confirmed cases in Spain, Italy and Germany from late February to mid August 2020. 
The \texttt{R} code for implementing the proposed model can be made available upon request. We also provide a Shiny web application \citep{shiny2020app} based on the model discussed in this paper at \url{https://ajalilian.shinyapps.io/shinyapp/}.

The structure of the paper is the following. 
The open data resources used in this study are introduced in Section~\ref{sec:data}. A model for the daily number of regional cases is considered in Section~\ref{sec:bymmodel}. As described in Section~\ref{sec:relrisk},  this model explains the spatio-temporal variations in the relative risk of each country in terms of a number of temporal, spatial and spatio-temporal random effects. The results of fitting the 
considered model to the number of daily confirmed cases in Spain, Italy and Germany are given in Section~\ref{sec:results}. The paper concludes in Section~\ref{sec:conclusion} with a few last remarks.

\section{Data on the daily number of COVID-19 cases}
\label{sec:data}

Governmental and non-governmental organizations across the world are collecting and reporting regional, national and global data on the daily number of confirmed cases, deaths and recovered patients and provide open data resources. Incompleteness, inconsistency, inaccuracy  and ambiguity of these open data are among limitations of any analysis, modeling and forecasting based on the data \citep{alamo2020open}. 
Particularly, the number of  cases mainly consist of cases  confirmed  by  a  laboratory  test and do not include infected asymptomatic cases and infected symptomatic cases without a positive laboratory test. 

In this study, we focus on the daily number of confirmed cases in Spain, Italy and Germany and  used the following open data resources.
\begin{description}
 \item[Spain:] DATADISTA, a Spanish digital communication medium that extracts the data on  confirmed cases (PCR and antibody test) registered by the Autonomous Communities of Spain and published by the Ministry of Health and the Carlos III Health Institute. DATADISTA makes the data available in an accessible format at the GitHub repository \url{https://github.com/datadista/datasets/tree/master/COVID%2019}. 
The daily accumulated number of total confirmed cases registered in the 19 Autonomous Communities of Spain are updated by DATADISTA on a daily bases. 
\item[Italy:] Data on the daily accumulated number of confirmed cases in the 20 regions of Italy are reported by the Civil Protection Department (Dipartimento della Protezione Civile), a  national  organization  in  Italy  that  deals  with  the  prediction, prevention and management of emergency events. These data are available at the GitHub repository \url{https://github.com/pcm-dpc/COVID-19} and are being constantly updated every day at 18:00.
\item[Germany:] The Robert Koch Institute, a federal government agency and research institute responsible for disease control and prevention, collects data and publishes official daily situation reports on  COVID-19 in Germany. Data on the daily accumulated number of confirmed cases in the 16 federal states of Germany extracted from the situation reports of the Robert Koch Institute are available at the GitHub repository \url{https://github.com/jgehrcke/covid-19-germany-gae} and are being updated on a daily basis. 
\end{description}

\begin{table}
\centering
\caption{Considered countries with their corresponding number of regions ($m$), length of study period and the estimated country-wide daily incidence rate ($\widehat{\varrho}_0$).}
\label{tab:country}
\renewcommand{\arraystretch}{1.3} 
\begin{tabular}{lcccc}
country & number of regions & study period & incidence rate \\\hline
Spain & 18$^{a}$ Autonomous Communities &  2020-02-25 to 2020-09-13 & $62.5\times10^{-6}$\\
Italy & 20 Regions &  2020-02-25 to 2020-09-15 & $24.0\times10^{-6}$ \\
Germany & 16 Federal States &  2020-03-03 to 2020-09-12 & $16.6\times10^{-6}$\\\hline
\end{tabular}
\footnotesize{$^a$ The Autonomous Communities Ceuta and  Melilla are merged into ``Ceuta y Melilla''}\\
\end{table}

\autoref{tab:country} summarizes the number of regions, study period and country-wide daily incidence rate of the data for each country.

Data on distribution population of the considered countries are extracted from the Gridded Population of the World, Version 4 (GPWv4), which provides estimates of the number of persons per pixel (1 degree resolution) for the year 2020 \citep{ciesin2018gridded}. These data are consistent with national censuses and population registers.

\section{Modeling daily regional counts}
\label{sec:bymmodel}

Suppose that a country, the spatial domain of interest, is partitioned into regions $A_{1},\ldots,A_m$, defined by administrative divisions such as states, provinces, counties, etc (see \autoref{tab:country}). 
Let $Y_{it}$ denote the number of new COVID-19 cases in region $A_i$, $i=1,\ldots,m$, at time (day) $t=1,\ldots,T$ and $E_{it}=\EE[Y_{it}]$ be the expected number of new COVID-19 cases in region $A_i$ at time $t$. 

\subsection{The null model of homogeneous incidence rates}

The expected number of new cases is given by $E_{it} = P_{i} \varrho_{it}$, where $P_{i}$ is the population of region $A_i$ and $\varrho_{it}$ is the incidence rate of COVID-19 in region $A_i$ at time $t$. Under the null model of spatial and temporal homogeneity of the incidence rate $\varrho_{it}=\varrho_0$ and 
\begin{equation}\label{eq:Ehat}
  \widehat{E}_{it} = P_i\widehat{\varrho}_0,
\end{equation}
provides an estimate for $E_{it}$, where 
\[
  \widehat{\varrho}_0 = \frac{1}{T} \sum_{t=1}^{T} \frac{\sum_{i=1}^{m} Y_{it}}{\sum_{i=1}^{m} P_i}
\]
is an estimate of the country-wide homogeneous daily incidence rate \citep{waller1997hierarchical}. The estimated daily incidence rate per million population ($10^{6}\widehat{\varrho}_0$) so far is around 68, 46 and 29 for Spain, Italy and Germany, respectively (see \autoref{tab:country}).

\subsection{Distribution of daily regional counts}

\cite{consul1973generalization} introduced 
a generalization of the Poisson distribution, which is a suitable model to most unimodal or reverse J-shaped counting distributions. Given nonnegative random rates $\Lambda_{it}$, $i=1,\ldots,m$, $t=1,\ldots,T$, we assume that $Y_{it}$'s are independent random variables following 
the generalized Poisson distribution  with
\[
  \PP(Y_{it}=y_{it}|\Lambda_{it},\varphi,\alpha) = 
 \exp\left(-\Omega_{it} - \Psi_{it} y_{it} \right)\frac{\Omega_{it}(\Omega_{it} + \Psi_{it} y_{it})^{y_{it} - 1}}{y_{it}!}, \quad y_{it}=0,1,\ldots,
\]
and the parameterization \citep{zamani2012functional}
\[
  \Omega_{it}=\frac{\Lambda_{it}}{1 + \varphi \Lambda_{it}^{\alpha-1}}, \quad 
  \Psi_{it} = \frac{\varphi \Lambda_{it}^{\alpha - 1}}{1 + \varphi \Lambda_{it}^{\alpha - 1}},
\]
where $\varphi\geq0$ and $\alpha\in \RR$. Then 
\[
   \PP(Y_{it}=y_{it}|\Lambda_{it},\varphi,\alpha) = \exp\left( -\frac{\Lambda_{it} + \varphi \Lambda_{it}^{\alpha-1} y_{it}}{1 + \varphi \Lambda_{it}^{\alpha-1}} \right) \frac{\Lambda_{it} \left( \Lambda_{it} + \varphi \Lambda_{it}^{\alpha-1} y_{it} \right)^{y_{it}-1}}{\left(1 + \varphi \Lambda_{it}^{\alpha-1} \right)^{y_{it}}y_{it}!}
\]
and
\[
  \EE[Y_{it}|\Lambda_{it},\varphi,\alpha] = \Lambda_{it}, \quad \Var(Y_{it}|\Lambda_{it},\varphi,\alpha) = \Lambda_{it} \left( 1 + \varphi \Lambda_{it}^{\alpha-1}\right)^{2}.
\]
Thus $\varphi$ is the dispersion parameter and the case $\varphi=0$ represents the ordinary Poisson  distribution (no dispersion) with 
\[
  \PP \left( Y_{it} = y_{it} | \Lambda_{it} \right) = \exp\left(-\Lambda_{it}\right) \frac{\Lambda_{it}^{y_{it}}}{y_{it}!}, \quad y_{it}=0,1,\ldots.
\]
Here, parameter $\alpha$ controls the shape (power) of the relation between the conditional variance of $Y_{it}|\Lambda_{it}$ and its conditional mean. For example, the relation between $\Var(Y_{it}|\Lambda_{it})$ and $\EE[Y_{it}|\Lambda_{it}]$ is linear if $\alpha=1$ and cubic if $\alpha=2$ \citep{zamani2012functional}.

\section{Modeling relative risks}
\label{sec:relrisk}

The underlying random rates $\Lambda_{it}$, $i=1,\ldots,m,t=1,\ldots,T$,  account for the extra variability (overdispersion), which may represent unmeasured confounders and model misspecification \citep{wakefield2007disease}. Variations of the random rate $\Lambda_{it}$ relative to the expected number of cases $E_{it}$ provide useful information about the spatio-temporal risk of COVID-19 in the whole spatial domain of interest during the study period.

\subsection{Relative risks}

In disease mapping literature, the nonnegative random quantities
\[
  \theta_{it} = \frac{\EE[ Y_{it} | \Lambda_{it}]}{\EE [Y_{it}]} = 
  \frac{\Lambda_{it}}{E_{it}}, \quad i=1,\ldots,m,
\]
are called the area-specific relative risks at time $t$ \citep[][Section~5.1.4]{lawson2018book}. 
Obviously $\EE\theta_{it}=1$ and 
\[
  \Cov( \theta_{it}, \theta_{i^{\prime}t^{\prime}}) =
  \frac{\Cov( \Lambda_{it}, \Lambda_{i^{\prime}t^{\prime}})}{E_{it}E_{i^{\prime}t^{\prime}}},
\]
which means that the temporal and spatial correlation structure of the underlying random rates $\Lambda_{it}$ determine the spatio-temporal correlations between $\theta_{it}$'s.  

By ignoring these correlations,  the standardized incidence ratio $\widehat{\theta}_{it}=Y_{it}/\widehat{E}_{it}$ provides a naive estimate for the relative risks \citep{lee2011comparison}. However, in a model-based approach the variations of the relative risks are often related to regional and/or temporal observed covariates and the correlation between $\theta_{it}$'s are explained in terms of regional and/or temporal random effects using, for example, a log linear model \citep{wakefield2007disease,lee2011comparison,lawson2018book}.

\subsection{A model for relative risks}

In the present study, we consider the log linear model
\begin{equation}\label{eq:thetamodel}
  \theta_{it} = \exp\left( \mu  + \beta d_{i} + \eta_{it} \right),
\end{equation}
where $\mu$ is the intercept and  $d_i$ is the population density of region $A_i$, i.e. the population of $A_i$, $P_i$, divided by the area of $A_i$. The population density is standardized to have mean  0 and variance 1 and $\beta$ is its regression coefficient. 
Moreover, $\eta_{it}$ is a zero mean random effect which represents spatio-temporal variations in relative risks due to temporal and spatial trend and correlation. Among many different possibilities, we assume that $\eta_{it}$ takes the additive form
\begin{equation}
 \eta_{it} = \delta_{t} +  \varepsilon_{t} + \zeta_{i} + \xi_i
\end{equation}
where $\delta_i$ represents the temporal trend, $\varepsilon_t$ accounts for temporal correlation and $\zeta_i$ and $\xi_i$ explain spatial correlation due to spatial distance and neighborhood relations among regions $A_1,\ldots,A_m$, respectively (see \autoref{tab:sptempterms}). 

\begin{table}
\centering
\caption{Considered terms in the additive model for the spatio-temporal random effect of the log-linear model for relative risks.}
\label{tab:sptempterms}
\begin{tabular}{clc}
 term & description & model \\\hline
 $\delta_t$ &  temporal trend & RW2 \\ 
 $\varepsilon_t$ & temporal correlation & AR(2) \\
 $\zeta_i$ & spatial correlation due to distance between regions & GMRF \\
 $\xi_i$ & spatial correlation due to neighborhood relation between regions & BYM \\\hline
\end{tabular}
\end{table}

The latent (stochastic) temporal trend $\delta_t$ is expected to be a smooth function of $t$. Since the second order random walk (RW2) model is appropriate for representing smooth curves \citep{fahrmeir2008identification},  $\boldsymbol{\delta}=(\delta_1,\ldots,\delta_{T})$ is assumed to follow 
a RW2 model, i.e.,
\[
  \Delta^{2}\delta_{t+1} = \delta_{t+1} - 2 \delta_{t} + \delta_{t-1} =\epsilon_{t}, \quad t=2, \ldots, T-1,
\]
where $\epsilon_2,\ldots,\epsilon_{T-1}$ are independent and identically distributed (i.i.d.)  zero mean Gaussian random variables with variance ${1}/{\tau_{\delta}}$. Here the precision parameter $\tau_{\delta}>0$ acts as a smoothing parameter enforcing small or allowing for large variations in $\delta_t$ \citep{fahrmeir2008identification}.

To account for temporal correlation, we assume that $\varepsilon_{t}$ follows a stationary autoregressive model of order 2, AR(2); i.e.,
\[
 \varepsilon_{t} = \psi_1 (1 - \psi_2) \varepsilon_{t-1} + \psi_2 \varepsilon_{t-2} + \epsilon^{\prime}_t, \quad t=2, \ldots, T,
\]
where $-1<\psi_1<1$ and $-1<\psi_2<1$ are the first and second partial autocorrelations of $\varepsilon_t$ and $\epsilon^{\prime}_2,\ldots,\epsilon^{\prime}_{T}$ are i.i.d. zero mean Gaussian random variables with variance  $1/\tau_{\varepsilon}$.

On the other hand, to account for spatial correlation, we assume that $\boldsymbol{\zeta}=(\zeta_1,\ldots,\zeta_{m})$ follows a Gaussian Markov random field (GMRF). More specifically, we assume that $\boldsymbol{\zeta}$ is a zero mean Gaussian random vector with the structured covariance matrix
\[
  \Var (\boldsymbol{\zeta}) = \frac{1}{\tau_{\zeta}} \left( \mathbf{I}_m - \frac{\omega}{e_{\max}} {\mathbf{C}}\right)^{-1}
\]
where $\mathbf{I}_{m}$ is the $m\times m$ identity matrix, $0\leq\omega<1$ and  $e_{\max}$ is the largest eigenvalue of the $m\times m$ symmetric positive definite matrix ${\mathbf{C}} = [{C}_{ii^{\prime}}]$. The entry $C_{ii^{\prime}}$ of matrix $\mathbf{C}$ represents to what extend the regions $A_i$ and $A_{i^{\prime}}$ are interconnected. For example, ${C}_{ii^{\prime}}$ can be related to a data on commuting or population movement between  regions $A_i$ and $A_{i^{\prime}}$. In absence of most recent and reliable movement data between the regions of Spain, Italy and Germany, we set $C_{ii^{\prime}}$ to be the Euclidean  distance between the centroids of $A_i$ and $A_{i^{\prime}}$.

In addition to interconnectivity and correlations due to spatial distance, the neighbourhood structure of regions $A_1,\ldots, A_m$ may induce spatial correlation among relative risks of regions because neighbouring regions often tend to have similar relative risks. To include spatial correlation due to neighborhood structure of regions in the model, we assume that $\boldsymbol{\xi}=(\xi_1,\ldots,\xi_m)$ follows a scaled version of the Besag-York-Molli{\'e} (BYM) model \citep{besag1991bayesian}; i.e., $\boldsymbol{\xi}$ is a zero mean Gaussian random vector with \citep{riebler2016intuitive}
\[
  \Var(\boldsymbol{\xi}) = \frac{1}{\tau_{\xi}} \left( (1-\phi) \mathbf{Q}^{-} + \phi \mathbf{I}_m \right).
\]
Here $\mathbf{Q}^{-}$ denotes the generalized inverse of the $m\times m$ spatial precision matrix  $\mathbf{Q}=[Q_{ii^{\prime}}]$ with entries 
\[
  Q_{ii^{\prime}} = \left\{ \begin{array}{lc} 
  n_i & i=i^{\prime} \\
  -1 & i\sim i^{\prime} \\
  0 & \text{otherwise}
\end{array}   \right.
\]
where $n_i$ is the number of neighbors of region $A_i$ and $i\sim i^{\prime}$ means that regions $A_{i}$ and $A_{i^{\prime}}$ share a common border. The parameter $\tau_{\xi}>0$ represents the marginal precision and $0\leq \phi\leq 1$ indicates the proportion of the marginal variance explained by the neighborhood structure of regions  \citep{riebler2016intuitive}.

\subsection{Prior specification and implementation}
\label{sec:priors}

In a Bayesian framework, it is necessary to specify prior distributions for all unknown parameters of the  considered model. 
The Gaussian prior with mean zero and variance $10^{6}$ is considered as a non-informative prior for the dispersion parameter of generalized Poisson distribution,  $\log\varphi$,  and for the parameters of the log linear model for the relative risks $\mu$, $\beta$, $\log\tau_{\delta}$, $\log\tau_{\varepsilon}$, $\log\tau_{\zeta}$, $\log\tau_{\xi}$, $\log\frac{\omega}{1-\omega}$, $\log\frac{\phi}{1-\phi}$, 
$\log\frac{1+\psi_1}{1-\psi_1}$ and $\log\frac{1+\psi_2}{1-\psi_2}$. 
The prior distribution for the $\alpha$ parameter of the generalized Poisson distribution is considered to be a Gaussian distribution with mean $1.5$ and variance $10^{6}$. 
\autoref{tab:modelpars} summarizes the model parameters and their necessary transformation for imposing the non-informative Gaussian priors.

\begin{table}
\centering
\caption{Parameters of the considered model for the daily number of new cases, their transformation for non-informative uniform (flat) priors and initial values.}
\label{tab:modelpars}
\renewcommand{\arraystretch}{1.3} 
\begin{tabular}{lccc}
parameter & notation &  transformation &  initial value \\\hline
dispersion parameter of generalized Poisson & $\varphi$ & $\log\varphi$ & 0 \\
shape parameter of of generalized Poisson & $\alpha$ & $\alpha$ & 1.5 \\
intercept & $\mu$ & $\mu$ & 0  \\
coefficient of population density & $\beta$ & $\beta$ & 0 \\
precision (smoothness) of the temporal trend & $\tau_{\delta}$ & $\log\tau_{\delta}$ & 0  \\
precision of $\varepsilon_t$ & $\tau_{\varepsilon}$  & $\log\tau_{\varepsilon}$ & 0 \\
first partial autocorrelation of $\varepsilon_t$  & $\psi_1$ & $\log\frac{1+\psi_1}{1 - \psi_1}$ & 0 \\
second partial autocorrelation of $\varepsilon_t$  & $\psi_2$ & $\log\frac{1+\psi_2}{1 - \psi_2}$ & 0 \\
precision of $\zeta_i$ & $\tau_{\zeta}$  & $\log\tau_{\zeta}$ &  0 \\
contribution of $\mathbf{C}$ in the variance of $\boldsymbol{\zeta}$  & $\omega$ & $\log\frac{\omega}{1-\omega}$ & 0 \\
precision of $\xi_i$ & $\tau_{\xi}$  & $\log\tau_{\xi}$ & 0 \\
contribution of $\mathbf{Q}^{-}$ in the variance of $\boldsymbol{\xi}$ & 
$\phi$ & $\log\frac{\phi}{1-\phi}$ & 0 \\\hline
\end{tabular}
\end{table}

Since all random effects of the model are Gaussian, the integrated nested Laplace approximation (INLA) method \citep{rue2009approximate} can be used for deterministic fast approximation of posterior probability distributions of the model parameters and latent random effects \citep{martins2013bayesian,lindgren2015bayesian}. 
The \texttt{R-INLA} package, an \texttt{R} interface to the \texttt{INLA} program and available at \url{www.r-inla.org}, is used for the implementation of the Bayesian computations in the present work. The \texttt{R} code can be made available upon request. 
The initial values for all parameters  in the \texttt{INLA} numerical computations are set to be the mean of their corresponding prior  distribution. The initial value of $\alpha$ is chosen to be one (see \autoref{tab:modelpars}).

\subsection{Bayesian model posterior predictive checks}
\label{sec:diagnostic}

For count data $Y_{it}$ and in a Bayesian framework, a probabilistic forecast is a posterior predictive  distribution on $\mathbb{Z}_{+}$. It is expected to generate values that are consistent with the observations (calibration) and concentrated around their means (sharpness) as much as possible \citep{czado2009predictive}. 
Following a leave-one-out cross-validation approach, let
\[
 \mathcal{B}_{-(it)} = \bigcap_{i^{\prime}\neq i=1}^{m} \bigcap_{t^{\prime}\neq t=1}^{T} \left\{Y_{i^{\prime}t^{\prime}}=y_{i^{\prime}t^{\prime}} \right\}
\]
be the event of observing all count values except the one for region $A_i$ at time $t$. 
\cite{dawid1984present}  proposed the cross-validated probability integral transform (PIT)
\[
  \mathrm{PIT}_{it}(y_{it}) = \mathbb{P}\left( Y_{it} \leq y_{it} \big| \mathcal{B}_{-(it)} \right)
\]
for calibration checks. Thus, $\mathrm{PIT}_{it}$ is simply the value that the predictive distribution function of $Y_{it}$ attains at the observation point $y_{it}$. 
The conditional predictive ordinate (CPO) 
\[
  \mathrm{CPO}_{it}(y_{it}) = \mathbb{P}\left( Y_{it} = y_{it} \big| \mathcal{B}_{-(it)}\right)
\]
is another Bayesian model diagnostic. Small values of $\mathrm{CPO}_{it}(y_{it})$ indicate possible outliers, high-leverage and influential observations \citep{pettit1990conditional}.


For count data, \cite{czado2009predictive} suggested
a nonrandomized yet uniform version of the PIT with 
\[
   F_{it}(u|y_{it}) = \left\{ 
   \begin{array}{ll}
    0, & u < \mathrm{PIT}_{it}(y_{it} - 1), \\
    \frac{u - \mathrm{PIT}_{it}(y_{it} - 1)}{\mathrm{PIT}_{it}(y_{it}) - \mathrm{PIT}_{it}(y_{it} - 1)}, \quad &  \mathrm{PIT}_{it}(y_{it} - 1) \leq u < \mathrm{PIT}_{it}(y_{it}), \\
    1, & u \geq \mathrm{PIT}_{it}(y_{it}),
   \end{array}
 \right.
\]
which is equivalent to
%
\begin{align*}
  F_{it}(u|y_{it}) &= \left( 1 - \frac{\mathrm{PIT}_{it}(y_{it}) - u}{ \mathrm{CPO}_{it}(y_{it})} \right) \I[ 0  < \mathrm{PIT}_{it}(y_{it}) - u \leq \mathrm{CPO}_{it}(y_{it})] \\ 
  &+  \I[\mathrm{PIT}_{it}(y_{it}) - u \leq 0],
\end{align*}
where $\I[\,\cdot\,]$ is the indicator function, 
because
\begin{align*}
  \mathrm{PIT}_{it}(y_{it} - 1) &= \mathbb{P}\left( Y_{it} \leq y_{it} - 1 \big| \mathcal{B}_{-(it)}\right) \\
  &= \mathbb{P}\left( Y_{it} \leq y_{it} \big| \mathcal{B}_{-(it)}\right) - \mathbb{P}\left( Y_{it} = y_{it} \big| \mathcal{B}_{-(it)}\right) \\
  &= \mathrm{PIT}_{it}(y_{it}) - \mathrm{CPO}_{it}(y_{it}).
\end{align*}
The mean PIT
\[
   \overline{F}(u) = \frac{1}{mT} \sum_{i=1}^{m}\sum_{t=1}^{T} F_{it}(u|y_{it}), \quad 0\leq u\leq 1,
\]
can then be comparing with the standard uniform distribution for calibration. For example, a histogram with heights
\[
  \overline{f}_j = \overline{F}\left(\frac{j}{J}\right) - \overline{F}\left(\frac{j-1}{J}\right), \quad j=1,\ldots,J,
\]
and equally spaced bins $\left[(j-1)/J,j/J\right)$, $j=1,\ldots,J$, can be compared with its counterpart from the standard uniform distribution with $f_j=1/J$.
Any departure from uniformity indicates forecast failures and model deficiencies. As mentioned in \cite{czado2009predictive}, U-shaped (reverse U-shaped)
histograms indicate underdispersed (overdispersed) predictive distributions and when central tendencies of the predictive distributions are biased, the histograms are skewed.

\section{Results}
\label{sec:results}

\autoref{tab:parest} presents the Bayesian estimates (posterior means) for every parameter of the considered model fitted to the daily number of new COVID-19 cases in Spain, Italy and Germany. The corresponding  95\% credible intervals of the model parameters are also reported in parentheses. 

\begin{table}
\centering
\caption{Posterior mean and 95\% credible  interval (in parentheses) for every parameter of the considered model fitted to the daily number of new COVID-19 cases in Spain, Italy and Germany.}
\label{tab:parest}
\renewcommand{\arraystretch}{1.5} 
\begin{tabular}{cccc}
parameter  & Spain  & Italy & Germany \\\hline
$\varphi$ & 0.613 (0.584, 0.645) & 1.196 (1.144, 1.251) & 0.836 (0.745, 0.910) \\
$\alpha$ & 1.482 (1.471, 1.492) & 1.319 (1.306, 1.330) & 1.301 (1.287, 1.314) \\
$\mu$ & -1.346 (-1.508, -1.205) & -0.951 (-1.101, -0.827) & -0.753 (-0.885, -0.620) \\
$\beta$ & -0.062 (-1.182, 1.065) & 0.085 (-0.096, 0.274) & 0.323 (0.031, 0.615) \\
$\tau_{\delta}$ &  0.062 (0.013, 0.127) & 0.440 (0.256, 0.668) & 0.200 (0.100, 0.374) \\
$\tau_{\varepsilon}$ & 57.9 (23.9, 126.1) & 54.9 (24.4, 101.3) & 7.11 (4.58, 11.45) \\
$\psi_1$ & 0.632 (0.573, 0.671) & 0.638 (0.605, 0.662) & 0.585 (0.524, 0.634) \\
$\psi_2$ & -0.932 (-0.981, -0.832) &  -0.973 (-0.998, -0.926) & -0.688 (-0.841, -0.416) \\
$\tau_{\zeta}$ &  0.185  (0.077, 0.352) & 17465.8 ( 1211.3, 67482.4) & 20175.4 (923.5, 72246.7) \\
$\omega$ & 0.337 (0.001, 0.963) & 0.315 (0.001, 0.958) & 0.302 (0.001, 0.962) \\
$\tau_{\xi}$ &  1087.5 (2.2, 6341.9) & 4.26 (2.13,   7.48) & 5.37 (2.32,  10.90) \\
$\phi$ &  0.422 (0.021, 0.958) & 0.876 (0.455,  0.999) & 0.622 (0.072, 0.986) \\\hline
\end{tabular}
\end{table}

Comparing the estimated parameters among different countries, it can be seen that the dispersion parameter $\varphi$ of the generalized Poisson distribution for Italy is higher than Spain and Germany, but its shape parameter $\alpha$ is around $1.5$ for the three countries, which implies that the variance of the daily counts in each region is approximately a quadratic function of their mean. The coefficient of the population density is not significantly different from zero for Spain and Italy, but is positive for Germany which indicates that regions with higher population density have larger relative risks. 

The precision parameters of the temporal random effects  imply that the temporal trend $\delta_t$ has at least 35 times larger contribution (smaller precision) than $\varepsilon_t$ which represents temporal correlation. The opposite signs of $\psi_1$ and $\psi_2$ indicate rough oscillations in $\varepsilon_t$. The spatial random effect $\zeta_i$ has larger contribution (smaller precision) than $\xi_i$ in the total variations of the relative risks only in Spain, while for Italy and Germany it is the opposite. This could be a result of large Euclidean distance between Spain continental European territory from two archipelagos territories, which is affecting the considered covariance structure of $\zeta_i$.

In summary, the higher contribution (lower precision) in the total variations of the relative risks for Spain, Italy and Germany  is due to the temporal trend, spatial correlation and finally temporal correlation, respectively. 
This may hint  that spatial correlations have a greater impact on the relative risks of COVID-19 than temporal correlations.

\begin{figure}
\centering
\includegraphics[width=0.95\textwidth]{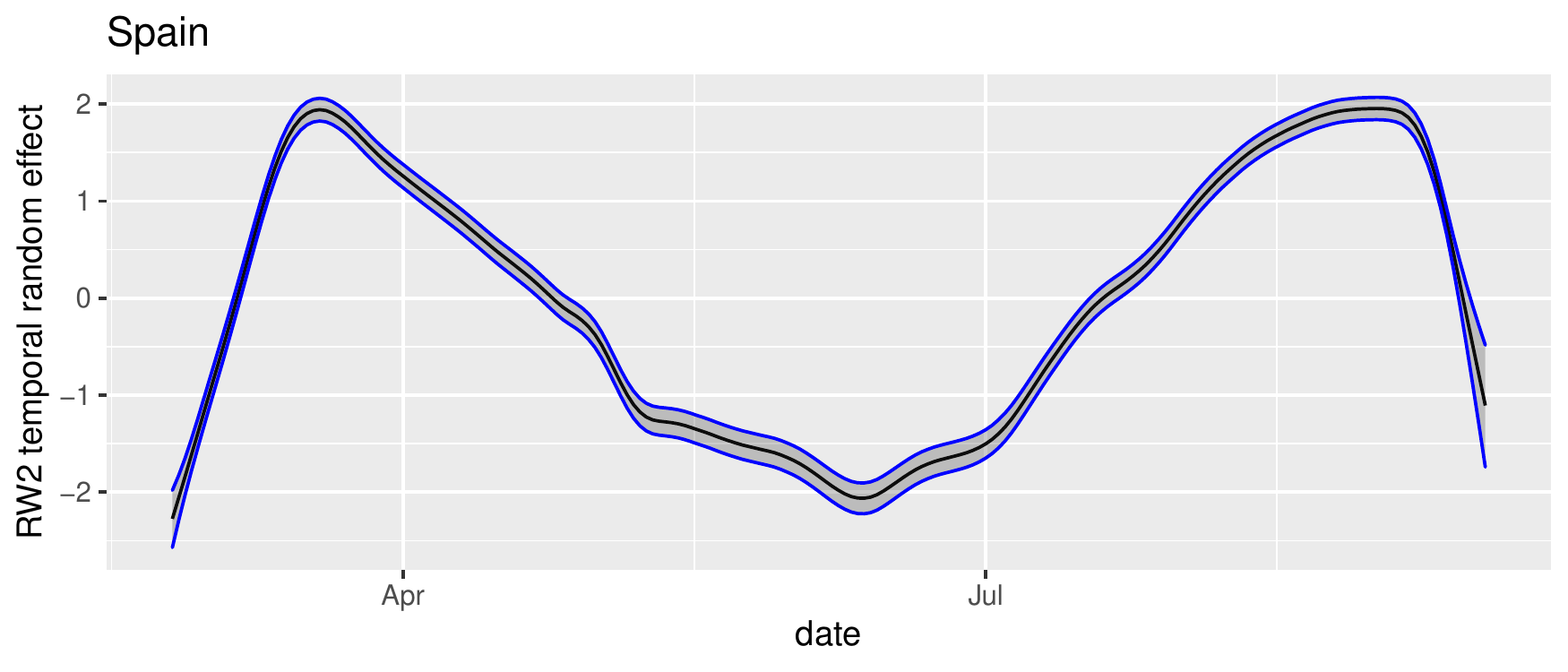}
\includegraphics[width=0.95\textwidth]{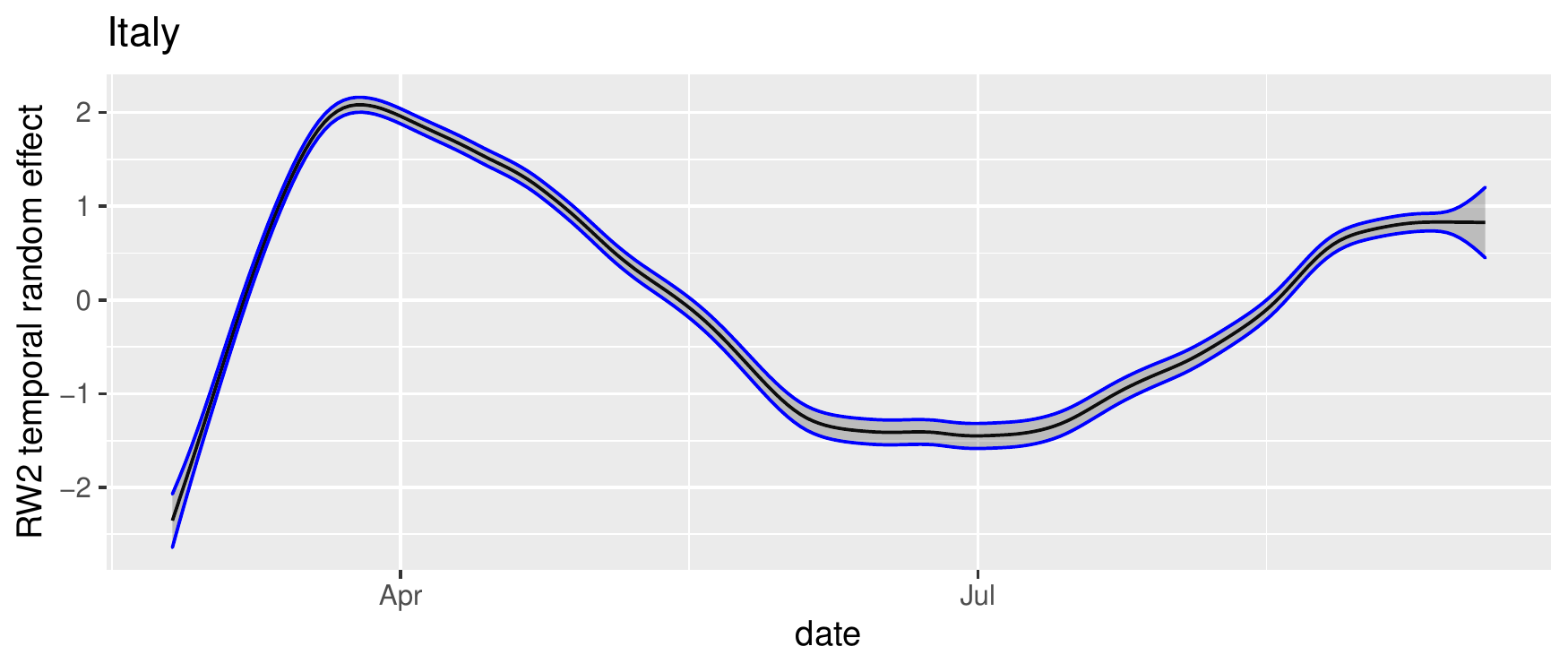}
\includegraphics[width=0.95\textwidth]{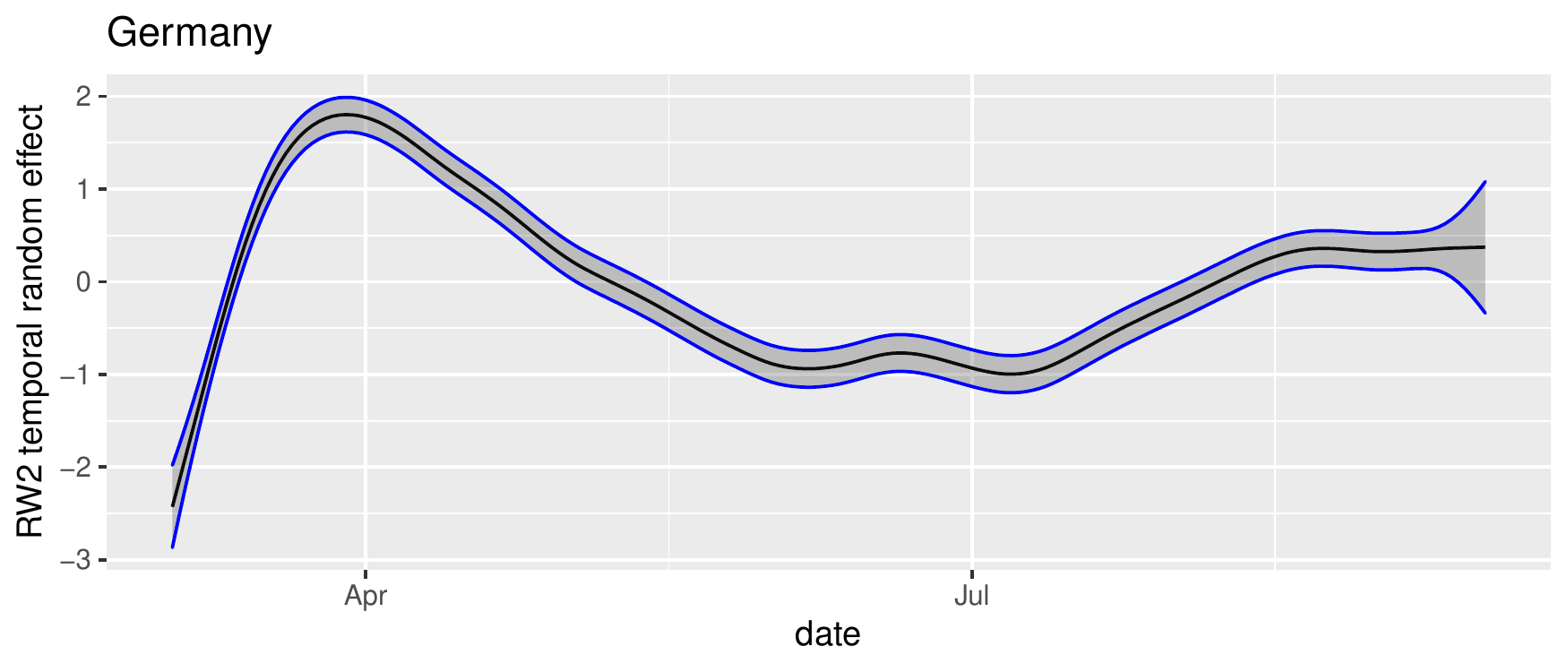}
\caption{Smoothed temporal trend of the relative risks of COVID-19, obtained from posterior mean and 95\% credible interval of the structured temporal random effect of the fitted model}
\label{fig:smoothtemp}
\end{figure}

The Bayesian estimates and 95\% credible intervals for the temporal trend $\delta_{t}$, $t=1,\ldots,T$, are shown in \autoref{fig:smoothtemp}. These plots can be interpreted as a smoothed temporal trend of the relative risk in the whole country. In fact,    \autoref{fig:smoothtemp} suggests that the COVID-19 epidemic in all three countries rapidly reached their peaks and slowly started to decline at the beginning of April and then increased and reached its maximum in August. In addition, the second wave of the epidemic seems to be stronger in Spain and Germany shows a more smoother trend during the study period.

\begin{figure}
\centering
\includegraphics[width=0.9\textwidth]{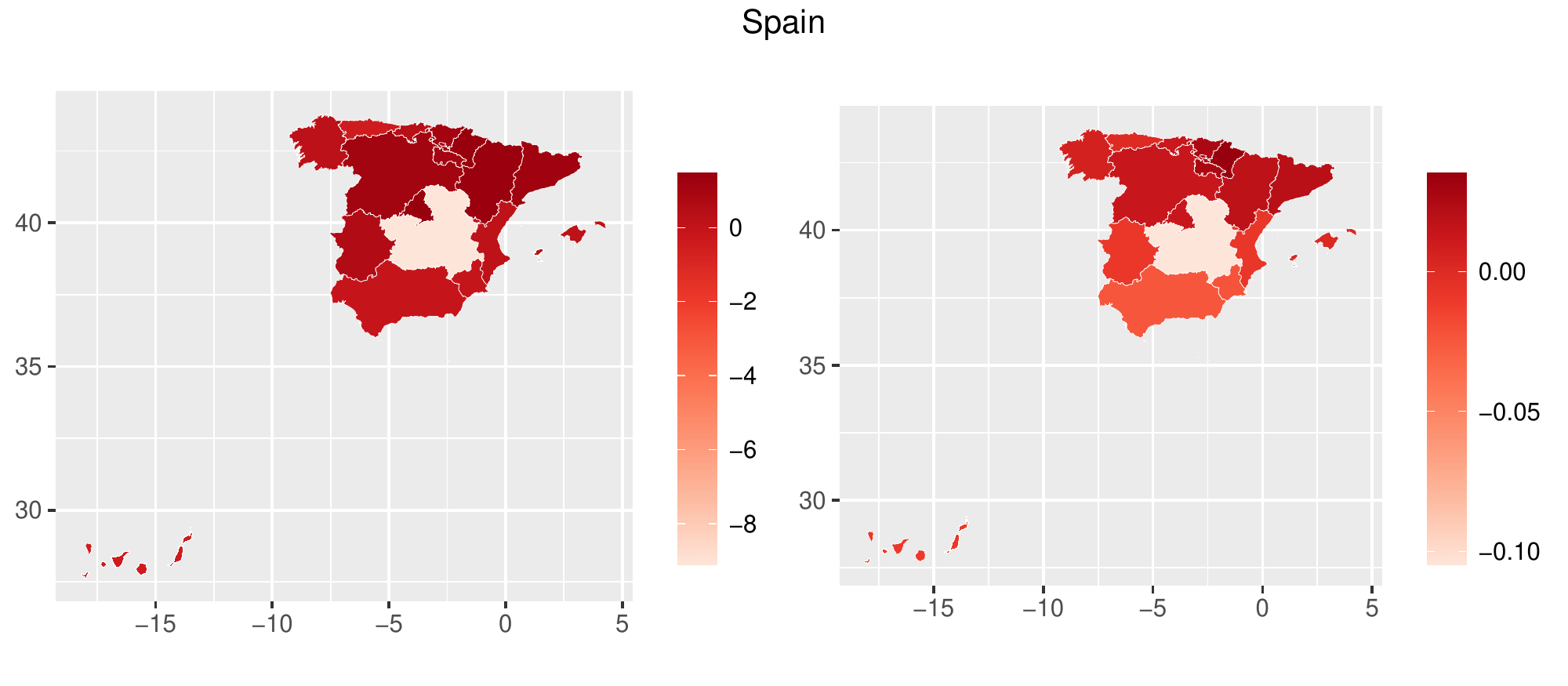}\\
\includegraphics[width=0.9\textwidth]{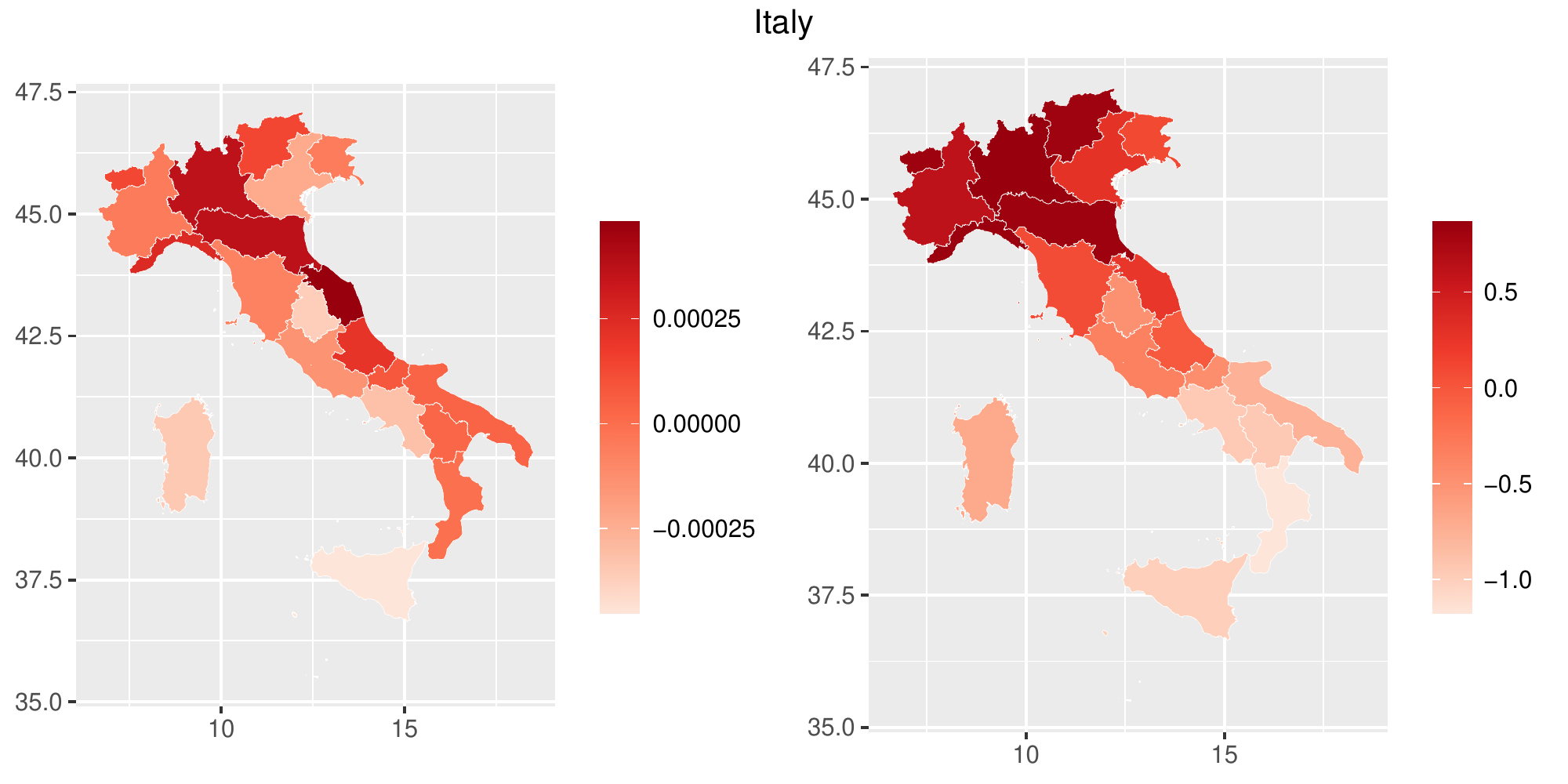}\\
\includegraphics[width=0.9\textwidth]{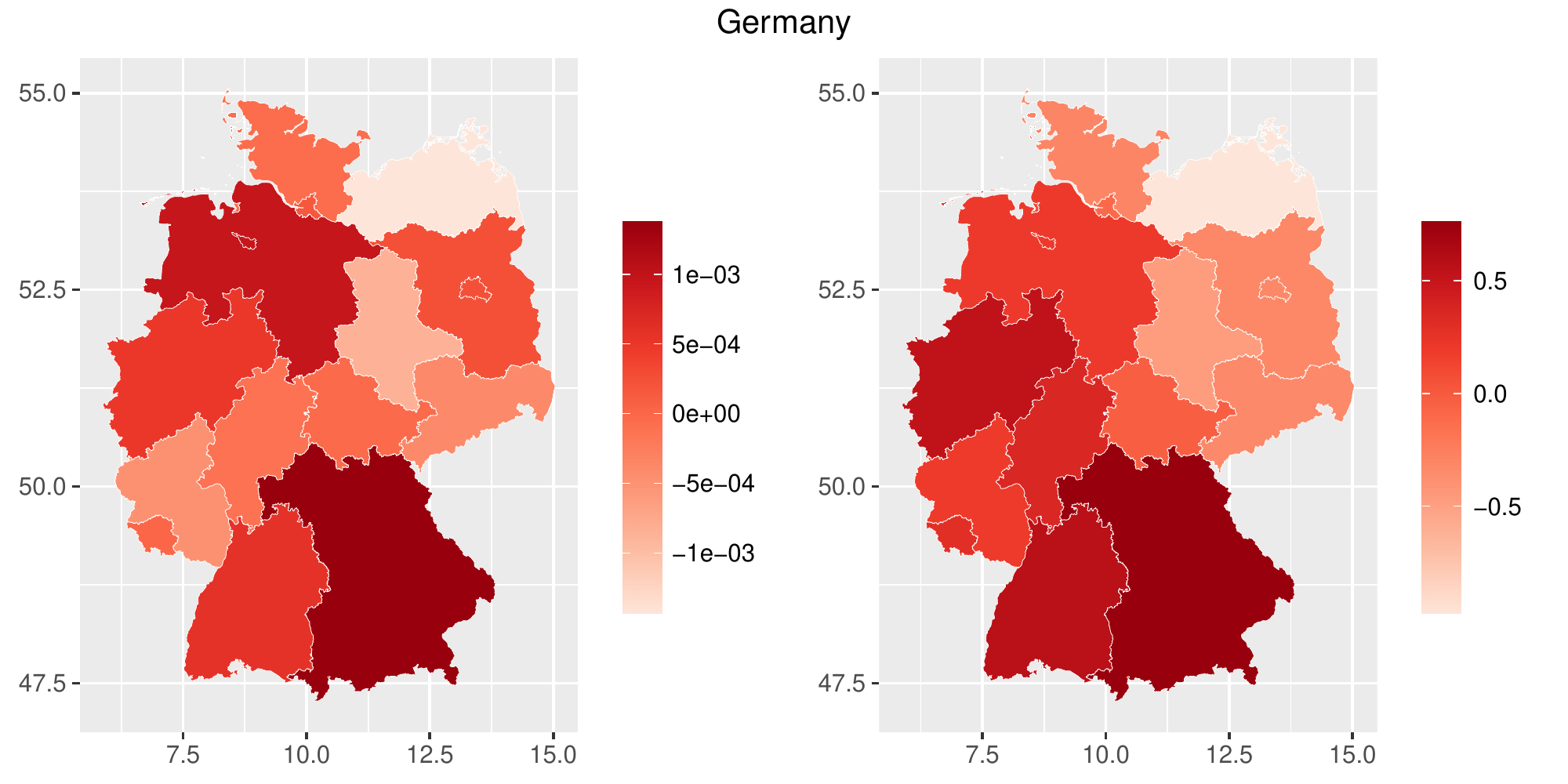}
\caption{Posterior mean of the spatial random effects $\zeta_i$ (left) and $\xi_i$ (right) in the fitted model}
\label{fig:spatialtrend}
\end{figure}

\autoref{fig:spatialtrend} shows the the posterior means of the spatial random effects $\zeta_{i}$ and $\xi_i$, $i=1,\ldots,m$, on the corresponding map of each country. The plot illustrates spatial heterogeneity of the relative risk of COVID-19 across regions in each country, particularly in Spain. Regions with positive (negative) $\zeta_{i} + \xi_i$ values are expected to have elevated (lower) relative risks than the the baseline country-wide risk during the study period.

\begin{figure}
\centering
\includegraphics[width=0.95\textwidth]{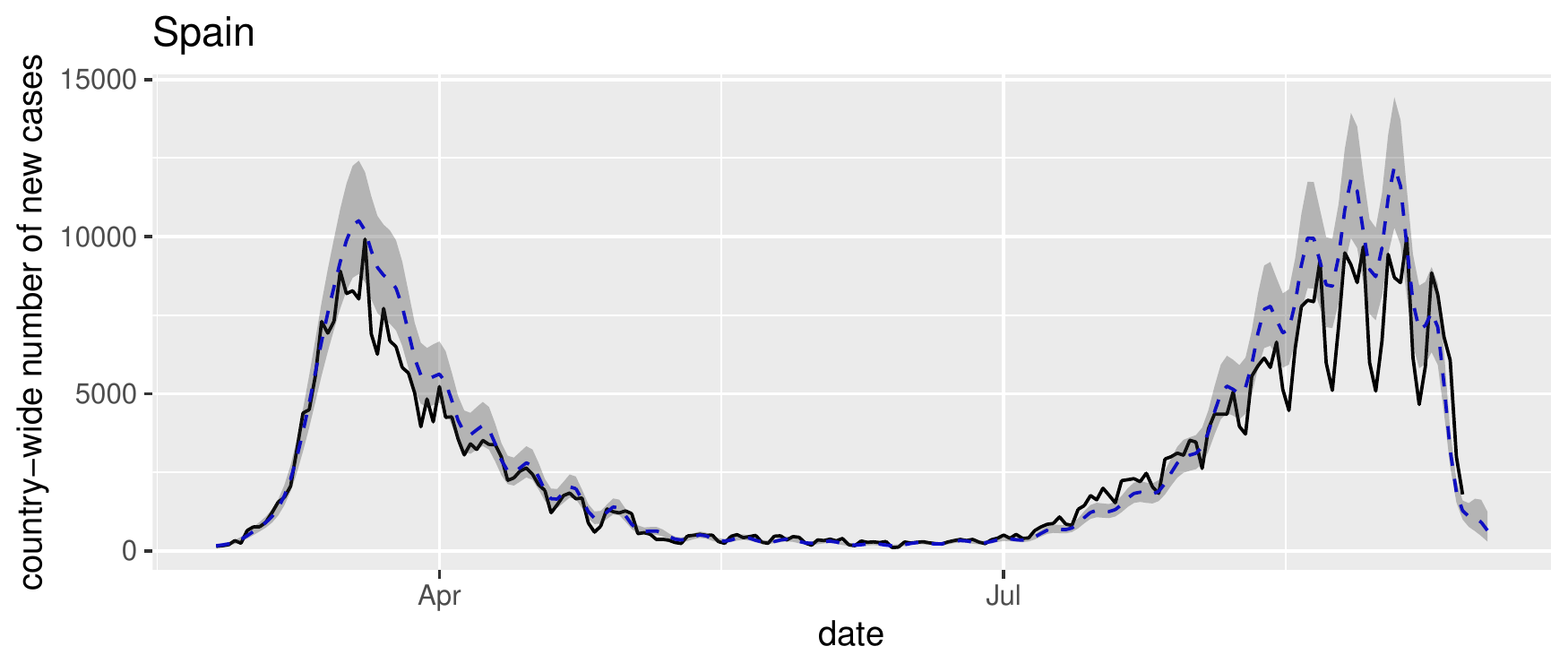}
\includegraphics[width=0.95\textwidth]{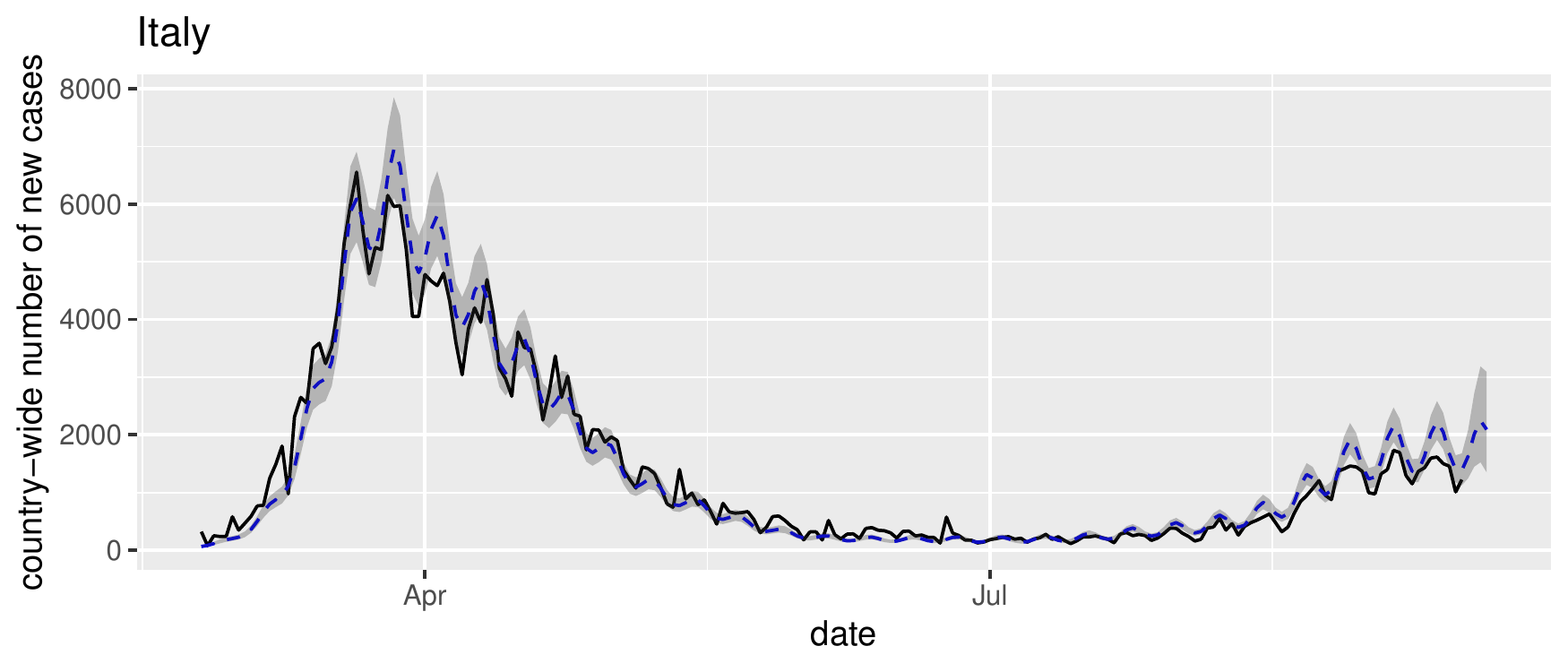}
\includegraphics[width=0.95\textwidth]{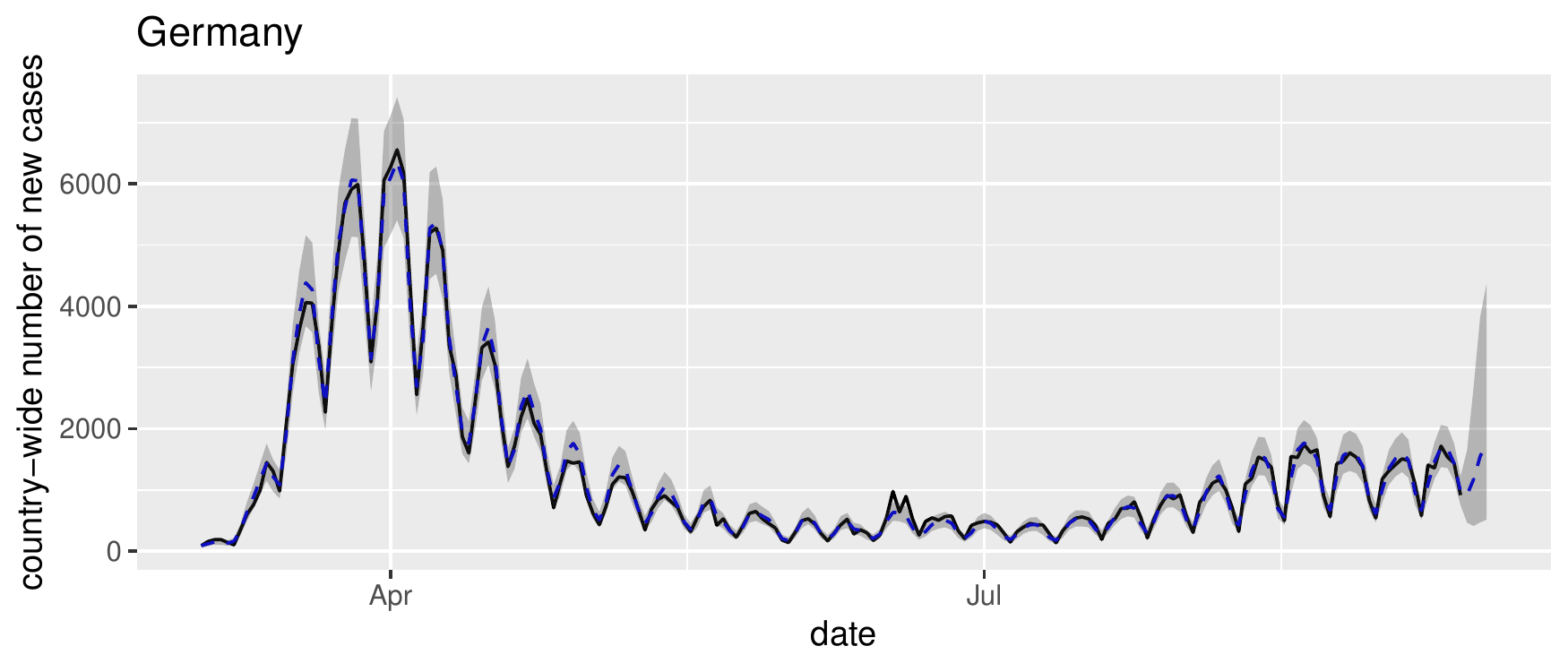}
\caption{Observed value (solid line), predicted value (dashed line) and 95\% prediction interval (grey area) for the daily number of new COVID-19 cases in the whole country, based on the posterior mean and 95\% credible interval of the spatially accumulated relative risks of the fitted model}
\label{fig:temptrend}
\end{figure}

In order to see how the estimated relative risks under the fitted model are in agreement with the observed data, \autoref{fig:temptrend} shows the spatially accumulated daily number of cases $\sum_{i=1}^{m}Y_{it}$, $t=1,\ldots, T$, and their expected values under the fitted model, namely the posterior mean  and 95\% credible interval of $\sum_{i=1}^{m}\widehat{E}_{it}\widehat{\theta}_{it}$, $t=1,\ldots, T$. Except some discrepancies for Spain and Italy,  the observed values are inside the 95\% credible intervals and close to the expected values under the fitted model. \autoref{fig:temptrend} in addition shows 4-days ahead forecasts of the total daily number of new cases at the end of study period of each country.

\begin{figure}
\centering
\includegraphics[width=0.85\textwidth]{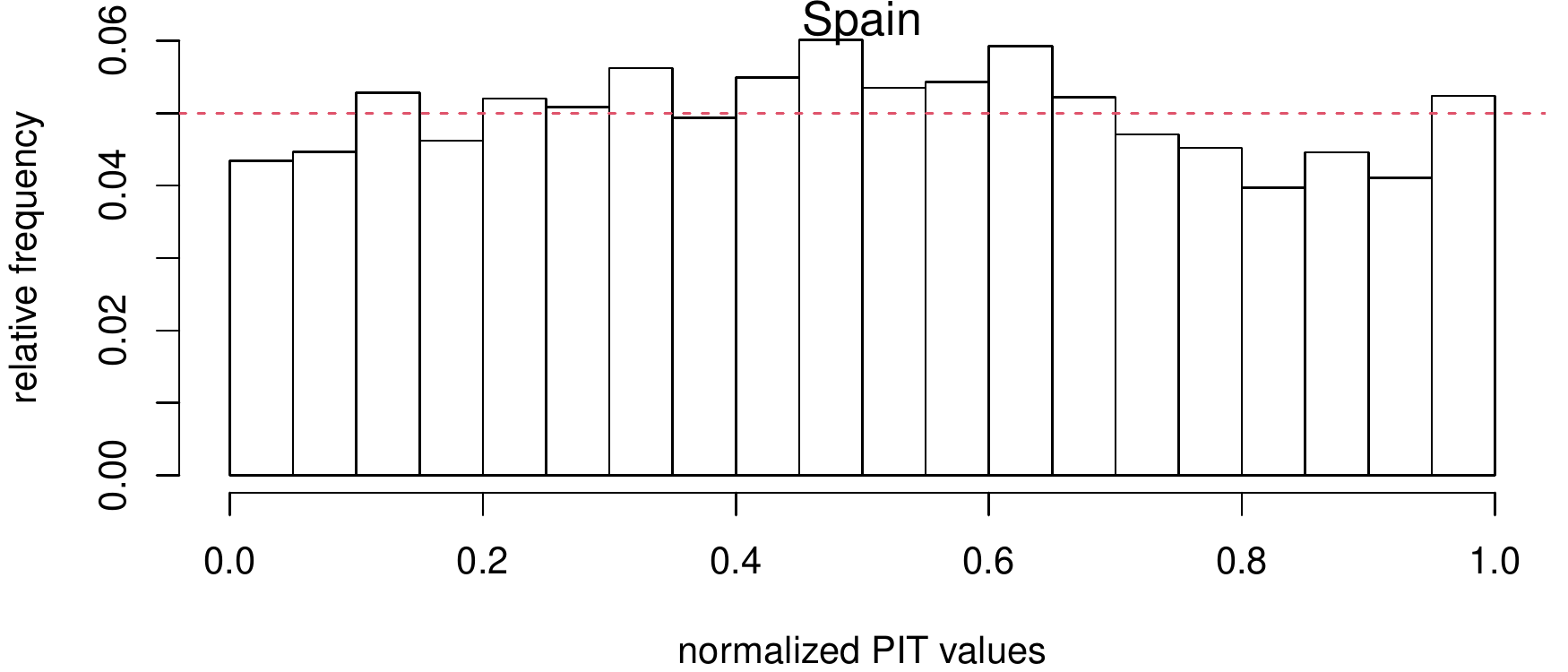}\vspace*{0.15cm}
\includegraphics[width=0.85\textwidth]{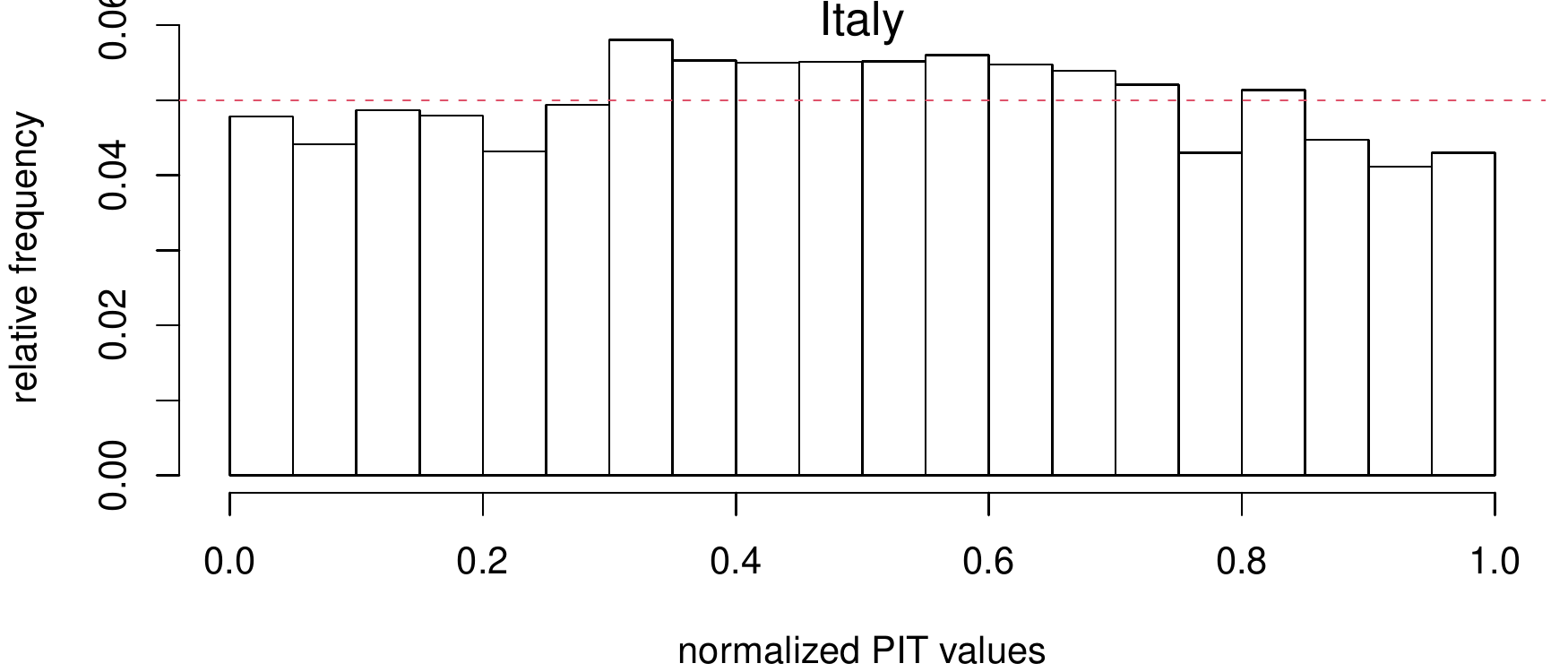}\vspace*{0.15cm}
\includegraphics[width=0.85\textwidth]{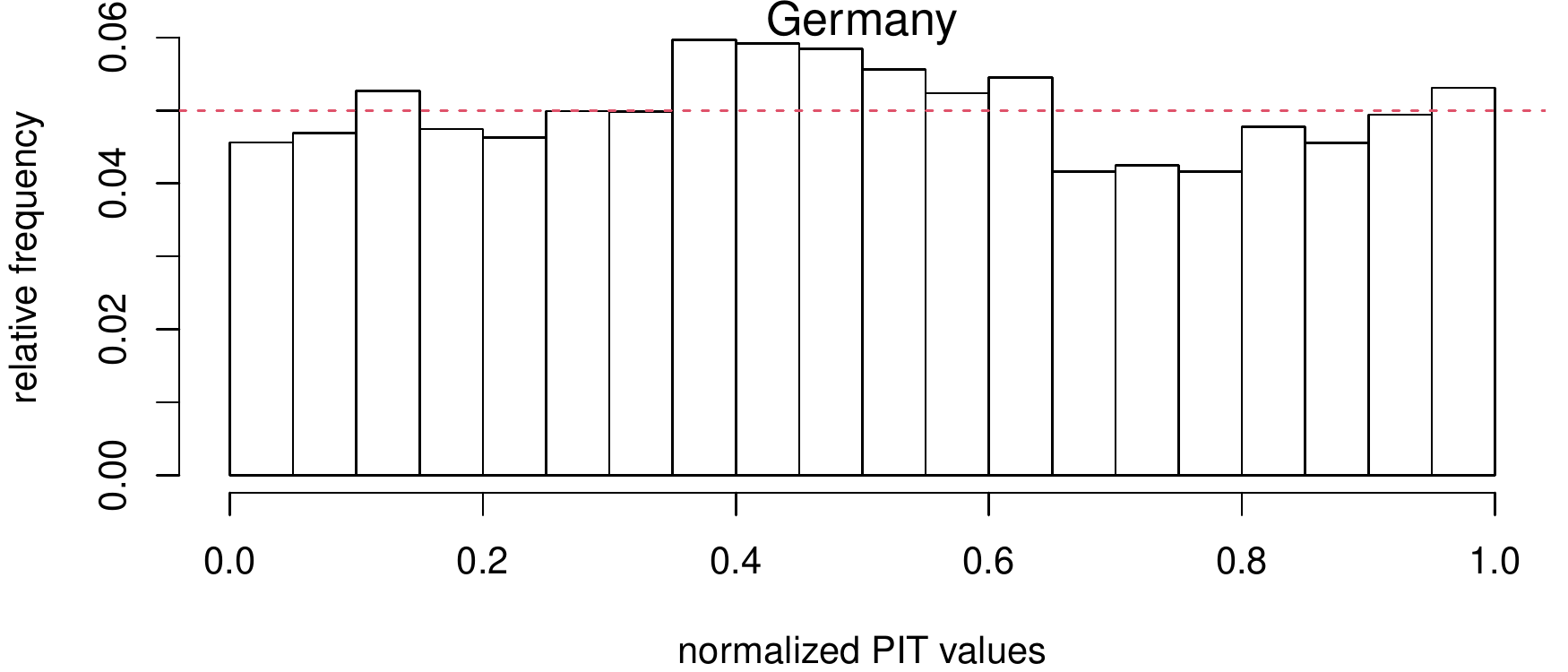}
\caption{Histograms of normalized PIT values to check for uniformity}
\label{fig:pit}
\end{figure}

Finally, histograms of the normalized PIT values described in Section~\ref{sec:diagnostic} are obtained using $J=20$ from the fitted models and plotted in \autoref{fig:pit}. The normalized PIT values for the fitted models to data do not show a clear visible pattern and the histograms seems to be close to the standard uniform distribution.

The above results and more details on observed and predicated values from the fitted model are also provided in an interactive Shiny web application at \url{https://ajalilian.shinyapps.io/shinyapp/}.

\section{Concluding remarks}
\label{sec:conclusion}

There are some limitations in the analyses and modeling of data on the number of new cases of COVID-19, including data incompleteness and inaccuracy, unavailability or inaccuracy of relevant  variables such as population movement and interaction, as well as the unknown  nature of the new COVID-19 virus. Nevertheless, understanding the underlying spatial and temporal dynamics of the spread of COVID-19 can result in detecting regional
or global trends and to further make informed and timely public health policies such as resource allocation.

In this study, we used a spatio-temporal model to explain the spatial and temporal variations of the relative risk of the disease in Spain, Italy and Germany. 
Despite data limitations and the complexity and uncertainty in the spread of COVID-19, 
the model was able to grasp the temporal and spatial trends in the data. However, the posterior predictive checks using the normalized probability integral transform (PIT) showed that there is room for the model improvements.

Obliviously, there are many relevant information and covariates that can be considered in our modeling framework and improve the model's predictive capabilities. One good possibility would be considering most recent and accurate human mobility amongst regions. We would expect our model would benefit from this information, which right now can not be accessed. Moreover, the considered spatio-temporal model in this paper is one instance among many possibilities. For example, one possibility is to include a random effect term in the model that represents variations due to joint spatio-temporal correlations; e.g., a separable sptaio-temporal covariance structure. However, the considered model was adequate and no joint spatio-temporal random effect term was considered to avoid increasing the model's complexity.

We focused here on a stochastic spatio-temporal model as a good alternative to existing deterministic compartmental models in epidemiology to explain the spatio-temporal dynamics in the spread of COVID-19. However, it should be emphasized that  one step forward would be  considering a combination of a deterministic compartmental model in terms of differential equations for the number of susceptible, exposed, infectious and  recovered cases with our sort of stochastic modeling approach. This is a clear novelty and a direction for  future research.

\section*{Acknowledgements}
J. Mateu has been partially funded by research projects UJI-B2018-04 (Universitat Jaume I),  AICO/2019/198 (Generalitat Valenciana) and MTM2016-78917-R (Ministry of Science).

\bibliographystyle{apalike}
\bibliography{library}


%
\section*{Conflict of interest}

 The authors declare that they have no conflict of interest.

\end{document}